\documentclass[showpacs,amsmath,amssymb,twocolumn,superscriptaddress]{revtex4}
\usepackage{graphicx}
\usepackage[all]{xy}
\usepackage{amsmath}
\usepackage{amssymb}
\usepackage{enumerate}
\newcommand{\be}{\begin{equation}}
\newcommand{\ee}{\end{equation}}
\newcommand{\ben}{\begin{eqnarray}}
\newcommand{\een}{\end{eqnarray}}
\newcommand{\bes}{\begin{subequations}}
\newcommand{\ees}{\end{subequations}}
\newcommand{\bb}{\bibitem}

\newcommand{\bfi}{\begin{figure}}
\newcommand{\efi}{\end{figure}}
\newcommand{\bc}{\begin{center}}
\newcommand{\ec}{\end{center}}
\newcommand{\sech}{\mbox{sech}}
\newcommand{\arccosh}{\mbox{arccosh}}

\begin{document}
\title{First-order formalism for flat  branes in generalized $N$-field models}

\author{D. Bazeia} 
\affiliation{Instituto de F\'\i sica Universidade de S\~ao Paulo, 05314-970 S\~ao Paulo SP, Brazil}
\affiliation{Departamento de F\'\i sica, Universidade Federal da Para\'\i ba, 58051-970 Jo\~ao Pessoa, PB, Brazil}
\affiliation{Departamento de F\'\i sica, Universidade Federal de Campina Grande, 58109-970 Campina Grande, PB, Brazil}

\author{ A. S. Lob\~ao Jr}
\affiliation{Departamento de F\'\i sica, Universidade Federal da Para\'\i ba, 58051-970 Jo\~ao Pessoa, PB, Brazil}
\email{}
\author{L. Losano}

\affiliation{Departamento de F\'\i sica, Universidade Federal da Para\'\i ba, 58051-970 Jo\~ao Pessoa, PB, Brazil}

\affiliation{Departamento de F\'\i sica, Universidade Federal de Campina Grande, 58109-970 Campina Grande, PB, Brazil}
\email{losano@fisica.ufpb.br}

\author{R. Menezes}
\affiliation{Departamento de Ci\^encias Exatas, Universidade Federal da Para\'\i ba, 58297-000 Rio Tinto, PB, Brazil.}

\affiliation{Departamento de F\'\i sica, Universidade Federal de Campina Grande, 58109-970 Campina Grande, PB, Brazil}

\email{rmenenezes@dce.ufpb.br}

\begin{abstract}
This work deals with braneworld scenarios obtained from $N$ real scalar fields, whose dynamics is generalized to include higher order power in the derivative of the fields.
For the scalar fields being driven by nonstandard dynamics, we show how a first-order formalism can be obtained for flat brane in the presence of several fields. We then illustrate
our findings investigating distinct potentials with one and two fields, obtaining stable standard and compact solutions in the braneworld theory. In particular, we have found different models describing the very same warp factor.
\end{abstract}


\maketitle


\section{introduction}

In the braneworld scenario with a single extra dimension of infinite extent, the braneworld consists in a domain wall embedded in the higher dimensional bulk. The defect represents the three-dimensional universe and, for more than one decade, distinct braneworld scenarios have been studied. In this environment, relevant issues which can be nicely discussed are, for instance, the gauge hierarchy and the cosmological constant problems \cite{Brane1,Brane2,Brane3,Brane4}. 

Although the original work \cite{Brane3} does not include scalar fields, models with one or more scalar fields coupled to gravity have been used to describe thick branes \cite{GW}. The spacetime around the brane can be five-dimensional anti-de Sitter (${\rm AdS}_5$) and, when the geometry inside the brane is Minkowski, it is called flat brane. However, in the case of four-dimensional anti-de Sitter (${\rm AdS}_4$) or de-Sitter (${\rm dS}_4$) geometry, we have a bent brane, which requires a nonvanishing cosmological constant. In this work we focus mainly on flat branes, thus we will only consider the case of vanishing cosmological constant.

The main features of these branes depend not only on the way the scalar fields couple to gravity, but also on how they self-interact and interact among themselves.
There are many studies which focus on standard dynamics, with the scalar fields interacting via the respective potential. The topological structures that arises from the scalar field constitutes a brane and, in this case, the main features of the brane only depend on the parameters introduced in the potential.

In recent years, however, one has studied different models, for which the dynamics is generalized to include higher order power on the derivative of the fields. These models were inspired by Cosmology, focusing mainly on dark energy \cite{KF1,KF2,KF3}. Other studies have been introduced recently \cite{A,GDT1,GDT2,GDT3,O,more,CGT}. In Ref.~\cite{GDT1}, for instance, one has found global defect structures:  kinks, global vortices and global monopolos.  In Ref.~\cite{GDT2},  some important aspects of kinks have been investigated, among them the conditions for the preservation of linear stability. Furthermore, in Ref.~\cite{GDT3} it was shown how the generalized models can support a first-order framework. As an interesting result, the generalized models may also support topological solutions with finite wavelength, being of compact nature \cite{RH}. In contrast with the standard kink, compactons only support massive states bounded to it \cite{GDT2,GDT3,BGM,BHM}. Another interesting result appears in the recent work \cite{T}, where one identifies the thick brane splitting caused by the spacetime torsion.

In this work, we focus on the flat brane scenario, with gravity being described standardly, but with the scalar fields being driven by nonstandard kinetic terms. The main aim is to introduce the first-order framework for several distinct scalar fields. For pedagogical reasons, we organize the work as follow. In Sec.~\ref{section2} we study generalized models describing flat branes in a five-dimensional bulk where the gravity is coupled with $N$ scalar fields. In Sec. \ref{sec:3} we focus on specific models, to illustrate how the main results work for one and two real scalar fields, with their corresponding solutions. We then move on to investigate stability in Sec.~\ref{stable}, and we conclude the work in Sec.~\ref{end} with some comments and conclusions.

\section{\label{section2}Generalized  braneworld models}

The models that we investigate describe five-dimensional gravity coupled to a set of $N$ scalar fields $\{\phi_1,\phi_2,\ldots,\phi_N\}$. They are driven by the following action
\be
{\cal S}=\int d^4 x dy \sqrt{|g|} \left(-\frac14 R + {\cal L}(\phi_i,X_{ij})\right),
\ee
where $i,j=1,2,\ldots,N$. Here we are using $4\pi G^{(5)}=1$ and $g=det(g_{ab})$, for $a,b=0,1,...,4$. We also define the quantities $X_{ij}$ as
\be
X_{ij}=\frac12\nabla \phi_i \nabla \phi_j ,
\ee
which are symmetric by construction. The line element for the five-dimensional spacetime can be written as 
\be
ds^2_5=g_{ab}dx^a dx^b = e^{2A}ds_4^2-dy^2.
\ee
Note that the four-dimensional spacetime is flat, so it has the following line element
\be
ds_4^2=dt^2-dx^2_1-dx_2^2-dx^2_3.
\ee
The function $A=A(y)$ controls the warp factor $e^{2A}$.  If the function $A(y)$ be even, the brane is symmetric. 

For the above metric, the Einstein equations are $G_{ab}=2\,T_{ab}$, with the energy-momentum tensor having the form
\be
T_{ab}=\nabla_{\!a} \phi_i \nabla_{\!b} \phi_j {\cal L}_{X_{ij}} - g_{ab} {\cal L} .
\ee
The equations of motion for the scalar fields are given by
\be
\nabla_a \left({\cal L}_{X_{ij}} \nabla^a \phi_j\right)={\cal L}_{\phi_i},
\ee
or more explicitly
\be
{\cal G}_{ij}^{ab}\nabla_a \nabla_b \phi_j + 2 X_{jl}{\cal L}_{X_{ij}\phi_l} - {\cal L}_{\phi_i}=0,
\ee
where ${\cal G}^{ab}$ has the form
\be
{\cal G}^{ab}_{ij}={\cal L}_{X_{ij}} g^{ab} + {\cal L}_{X_{il}X_{jm}} \nabla^a \phi_l \nabla^b \phi_m.
\ee
Here we are using the notation: ${\cal L}_{X_{ij}}=\partial {\cal L}/\partial X_{ij}$ and ${\cal L}_{\phi_i}=\partial {\cal L}/\partial \phi_i$, etc. 

As usual, we suppose that the scalar fields are static, and also, it only depends on the extra dimension. Therefore, we have $A=A(y)$ and $\phi_i=\phi_i(y)$ and the $N$ equations of motion for scalar fields reduce to
\be\label{fieldeqs}
({\cal L}_{X_{ij}}\!+2 {\cal L}_{X_{il}X_{jm}} X_{lm}) \phi_j^{\prime\prime} -2 X_{jl} {\cal L}_{X_{ij}\phi_l} +{\cal L}_{\phi_i}\!=\!-4{\cal L}_{X_{ij}} \!\phi_j^\prime \!A^\prime,
\ee
where prime denote derivative with respect to the extra dimension and $X_{ij}=-\phi_i^\prime\phi^\prime_j/2$.  

An important characteristic of the brane, the energy density
\be
\label{Klaus}
\rho =T_{00} = - e^{2A(y)} {\cal L},
\ee 
can be found explicitly, for the models to be investigated below.

We take the standard case as 
\be
\label{SC}
{\cal L}=X - V(\phi_i),
\ee 
where 
\be\label{hij}
X=h_{ij}(\phi_l)X_{ij},
\ee
with $h_{ij}(\phi_l)$ being a symmetric matrix that represents the metric on the scalar target space \cite{DeWolfe:1999cp}. 

In the case of flat brane we get the Eisntein equations
\bes\label{Eis0}
\ben
A^{\prime\prime}&=&\frac{4}{3}X_{ij}{\cal L}_{X_{ij}} \label{Appij},\\
A^{\prime2} &=& \frac13 \left({\cal L}-2X_{ij}{\cal L}_{X_{ij}}\right) \label{Apij}.
\een
\ees
The second equation is the null energy condition. This impose that the brane pressure is aways positive. Thus, the scalar fields model must obey the condition
${\cal L} - 2X_{ij}{\cal L}_{X_{ij}}>0$.  If we now multiply each one of the equations in \eqref{fieldeqs} by the corresponding $\phi_i$ and add them, we get 
\be\label{eqofmotionp}
({\cal L}-2X_{ij}{\cal L}_{X_{ij}})^{\prime}=8 A^\prime X_{ij}{\cal L}_{X_{ij}}.
\ee 
If we substitute the Eq. \eqref{Apij} above, we recover Eq. \eqref{Appij}. Therefore, the equations \eqref{Eis0} are not independent from each other. Also, we recall that \eqref{eqofmotionp} can be obtained from the Bianchi identity $\nabla^a G_{ab}=0$.

We choose the derivative of the warp factor with respect to the extra dimension to be a function of the $N$ scalar fields, in the form 
\be\label{aprime}
A^\prime=-\frac13 W(\phi_i),
\ee
where $W=W(\phi_i)$ is a function of the $N$ scalar fields $\phi_i$. Substituting this into \eqref{Appij}, we can write
\be
 W_{\phi_i} \phi^{\prime}_i=2 \phi_i^\prime \phi^\prime_j {\cal L}_{X_{ij}} .
\ee
A possible set of solutions for this equation is
\bes\label{philphi}
\be
 \phi^\prime_j {\cal L}_{X_{ij}} =\frac12 W_{\phi_i} 
\ee
which is the same equation that appears in the absence of gravity \cite{GDT3}.  Note that the equation \eqref{Apij} leads to the constraint 
\be\label{cons}
{\cal L} - 2 X_{ij} {\cal L}_{X_{ij}} =\frac13 W^2
\ee 
\ees
Using the above $N+1$ equations \eqref{philphi}, we can show that the derivative of the fields can be expressed in terms of the fields themselves, that is,
\be\label{system}
\phi_i^\prime = \phi_i^\prime(\phi_j).
\ee 
It is not difficult to show that for fields that obey the constraint \eqref{cons}, the equations \eqref{system} solve the second-order equations \eqref{fieldeqs} and \eqref{Appij}. Also, we can write the energy density as
\be 
\rho= e^{2A(y)} \left[\frac12\frac{dW}{dy}-\frac13 W^2\right].
\ee

In this paper, we focus attention on models described by ${\cal L}={\cal L}(\phi_i,X)$,
with $X$ defined by Eq.~\eqref{hij}. In this case, we can rewrite the equations \eqref{philphi} in the form
\bes\label{REDe}
\ben
\phi^\prime_i {\cal L}_{X}&=& \frac12 h^{-1}_{ij}(\phi_l)W_{\phi_j} \label{akdapo}\\
{\cal L}-2X{\cal L}_X &=& \frac13 W^2\label{alkl}
\een
\ees
where $h^{-1}_{ij} h_{jl}=\delta_{il}$. We can use Eq. \eqref{akdapo} to write
\be\label{XLX2}
X {\cal L}_X^2=-\frac18 h^{-1}_{ij}(\phi_l) W_{\phi_i} W_{\phi_j}
\ee
which is a very useful expression, to be used in the calculations that follow below. 

As an illustration, let us consider  the standard case given by Eq. \eqref{SC}.
We use the Eq \eqref{akdapo} to obtain the set of first order equations
\bes
\be 
\phi^\prime_i= \frac12 h^{-1}_{ij}(\phi_l)W_{\phi_j}.
\ee
We then use Eq. \eqref{XLX2} to write $X=-\frac18 h^{-1}_{ij}(\phi_l) W_{\phi_i} W_{\phi_j}$. Now, substituting this equation into Eq.~\eqref{alkl}, we find the potential
\be\label{standardcaseV}
V(\phi_i)=\frac18 h^{-1}_{ij}(\phi_i) W_{\phi_i} W_{\phi_j} - \frac13 W^2.
\ee
\ees
We note that this is one of the results of Refs.~{\cite{DeWolfe:1999cp}}.

As a second example, let us introduce the kinematically modified case for $N$ fields, given by  
\begin{equation}\label{L.modifi}
{\cal L}=\frac{2^{n-1}}{n}X|X|^{n-1}-V,
\end{equation}
where $n=1,2,3,...$. This generalizes the model with one field, introduced in Ref.~\cite{BGM}. The first-order equations are
\bes
\be\label{eqnova}
\phi_i^\prime=\frac12 h^{-1}_{ij}(\phi_l)W_{\phi_j}\left[\frac14 h_{ks}(\phi_l)W_{\phi_k}W_{\phi_s}\right]^{\frac{1-n}{2n-1}}.
\ee
We use the Eq.~\eqref{XLX2} in order to obtain the relation: $2^{2n+1} X|X|^{2(n-1)} = - h^{-1}_{ij}(\phi_l)W_{\phi_i} W_{\phi_j}$. With this and the Eq.~\eqref{alkl}, we can write the potential
\ben\label{eqpotenti}
V(\phi_i)=\frac{2n-1}{2n}\left(\frac14 h^{-1}_{ij}(\phi_l)W_{\phi_i} W_{\phi_j}\right)^{\frac{n}{2n-1}}\!\!\!-\frac13 W^2.\;\;
\een
\ees

Comparing this expression with Eq. \eqref{standardcaseV} for the potential of the standard case, we see that only the $W_{\phi_i}$ portion
of the potential is changed. There is a simple reason for this: the $W$ portion of the potential is the geometric contribution,
and the gravity portion of the model remains unchanged.

\section{Specific models}\label{sec:3}

Let us now consider explicit examples of scalar field models, described by one and by two real scalar fields.

\subsection{One-field models}

Let us first study models described by a single scalar field. Here, however, we change the standard strategy, fixing the profile of the solution
and then finding the respective $W$. For a single field, the equation \eqref{eqnova} and the potential \eqref{standardcaseV} change to 
\bes
\ben\label{eqnova2}
\phi^\prime&=&\left(\frac12 W_{\phi}\right)^{\frac{1}{2n-1}},
\\
V(\phi)&=&\frac{2n-1}{2n}\left(\frac12 W_{\phi}\right)^{\frac{2n}{2n-1}}\!\!\!-\frac13 W^2\label{pot28},
\een
\ees
with $h=1$, evidently. For $W_\phi \neq 0$, the warp factor can be expressed in the form
\be
A(\phi(y))=-\frac13 \int d\phi  \left(\frac{2}W_{\phi}\right)^{\frac{1}{2n-1}}W,
\ee
and the energy density as
\be
\rho(\phi(y))=\left[\left(\frac{W_{\phi}}2 \right)^{\frac{2N}{2n-1}}-\frac13 W^2\right]e^{2A(y)}.
\ee

\begin{figure}[h!] 
\includegraphics[scale=0.45]{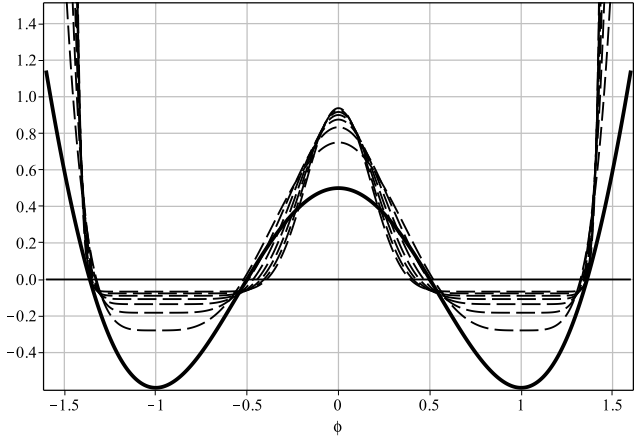} 
\caption{\label{figure1}Profile of the potential $V(\phi)$ given by Eq.~\eqref{pot28}, for $W(\phi)$ as in Eq.~\eqref{W_tanh}, for $n=1$ (solid line) and for $n=2,3,4,5,6,7,8$ (dashed lines).}
\end{figure}

We choose the kink solution 
\begin{equation}\label{solkink}
\phi(y)=\tanh(y).
\end{equation}
In the absence of gravity, this is the well known solution of the $\phi^4$ model, with spontaneous symmetry breaking. From the solution, using the  Eq.~\eqref{eqnova2}, we can reconstruct $W(\phi)$. In this case, we get to
\ben
W(\phi)&=&2\int \!d\phi\, (1-\phi^2)^{2n-1}, \nonumber \\
&=&2\phi \times \,{_2}F_1\left(\frac12,-2n+1;\,\frac32;\,\phi^{2}\right).\label{W_tanh}
\een
This hipergeometric function ${_2}F_1$ is a polynomial with degree $4n-1$. For example,  we can write explicitly   
\bes
\ben
W(\phi)&=&2\phi-\frac23\phi^3,\\
W(\phi)&=&2\phi-2\phi^3+\frac65\phi^5-\frac27\phi^7,
\een
\ees
for $n=1$ and $n=2$, respectively. Using the Eq. \eqref{pot28}, we can write the following potentials
\bes
\ben
V(\phi)&=&\frac12\left(1-\phi^2\right)^2-\frac43\left(\phi-\frac13\phi^3\right)^2,\\
V(\phi)&=&\frac34\left(1-\phi^2\right)^4-\frac43\left(\!\phi\!-\!\phi^3\!+\!\frac35\phi^5\!-\!\frac17\phi^7\!\right)^2\!,\;\;\;\;
\een
\ees
for $n=1$ and $n=2$, respectively. In general, the potential is a polynomial with degree $8n-2$. The potential has five local extrema: $\phi_0$, being the central extremum, is a local maximum at $\phi_0=0$, where $V(\phi_0)=(2n-1)/(2n)$ is always a positive constant. Two other local extrema are local minima: they are $\phi_{\pm}=\pm 1$, with
\be
V(\phi_{\pm})=-\frac{\pi}3 \left(\frac{ \Gamma(2n)}{\Gamma(2n+\frac12)}\right)^2,
\ee
which is negative. There are two other local maxima. The behavior of the potential around the central maximum is shown in Fig. \ref{figure1}. Moreover, we can note that
$V(\phi\to\pm\infty) \to -\infty$; thus, although it is not shown in the respective figures, the potentials go asymptotically to $-\infty$. 

\begin{figure}[h!] 
\centering
\includegraphics[scale=0.45]{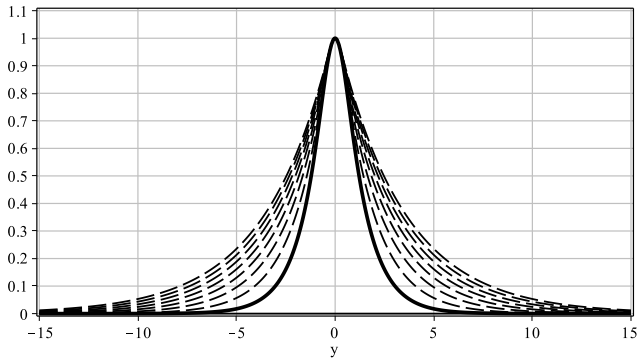} 
\caption{\label{figure2}Profile of the warp factor $e^{2A(y)}$, for $n=1$ (solid line) and for $n=2,3,4,5,6,7,8$ (dashed lines).}
\end{figure}

Despite having the same solution, these models have other distinct characteristics, as the warp function and the energy density. The warp functions are
\bes
\ben
A(y)\!&=&\!\frac49\ln S+\frac{S^2}{9}-\frac1{9}\label{A1igual},\\
A(y)\!\!&=&\!\!\frac{32}{105}\ln S+\frac{S^2}{105}+\frac{S^4}{35}+\frac{S^6}{63}-\frac{38}{315},
\een
\ees
for $n=1$ and $n=2$, respectvely, where $S=\sech(y)$. Here we fix $A(0)=0$. In Fig.~\ref{figure2} we depict the warp function for some values of $n$. We note that the warp factor $e^{2A}$ decays slower for bigger $n$. This can be verified from the behavior of the warp factor far outside the brane:
\ben
A_\infty(y) &\to& -\frac{W(\phi_+)}{3} |y|  
= -\frac{\sqrt{\pi}}{3} \frac{\Gamma(2n)}{\Gamma(2n+\frac12)} |y| .
\een
The ratio between the Gamma functions goes to zero for increasing $n$.

\begin{figure}[h!] 
\centering
\includegraphics[scale=0.45]{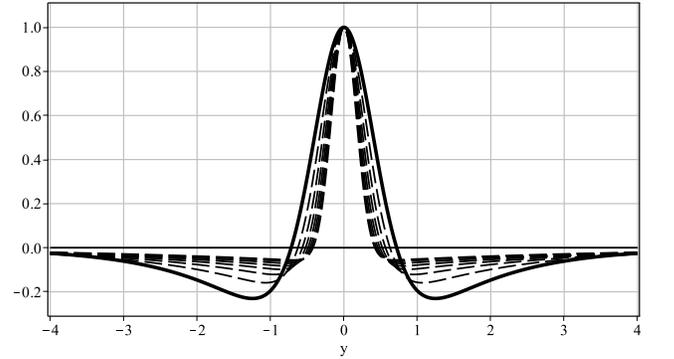} 
\caption{\label{figure3}Profile of the energy density $\rho(y)$ , for $n=1$ (solid line) and $n=2,3,4,5,6,7,8$ (dashed lines).}
\end{figure}

The energy densities also depend on $n$. For $n=1$ and $2$, the expressions are
 \ben
\rho(y)&=&h_1(y)\exp\left(\frac{2S}9 \right),\\
\rho(y)\!\!&=&\!\!h_2(y)\exp\!\left(\frac{16 S}{105}\!+\!\frac{2 S^4}{35}\!+\!
\frac{2 S^6}{63}\!\right),
\een
where 
\ben h_1(y)&=&\frac{S^{\frac89}}{27e^{\frac29}}\left[4S^6+39S^4-16\right].\\
h_2(y)&=&\frac{S^{\frac{32}{105}}}{3675e^{\frac{76}{315}}} \Big[100S^{14}+ 140 S^{12}+ 224 S^{10} \nonumber\\&&+4235S^8  -1024\Big].
\een
In the Fig. \ref{figure3} we depict the profile of the energy density for some values of $n$.    

\begin{figure}[h!] 
\includegraphics[scale=0.45]{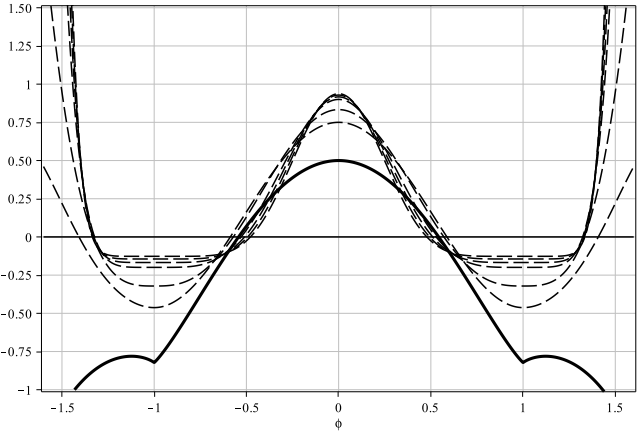} 
\caption{\label{figure4}Profile of the potential $V(\phi)$ \eqref{pot28} for $W(\phi)$ given by Eq.~\eqref{W_compact}, for $n=1$ (solid line) and for $n=2,3,4,5,6,7,8$ (dashed lines).}
\end{figure}

Another possibility is to take the one field model that support compact solution. We choose the specific form
\begin{equation}\label{compact}
\phi(y) = 
\begin{cases} 
{\rm sgn }(y) &  \mbox{ for } |y|>\frac{\pi}{2}\\ 
\sin(y) &  \mbox{ for } |y|\leq\frac{\pi}{2}
\end{cases}
\end{equation}
This solution obeys the first order equation $\phi^\prime = \sqrt{|1-\phi^2|}$. Thus, using the Eq.~\eqref{eqnova2}, we can reconstruct $W(\phi)$ as
\ben
W(\phi)\!\!&=&\!\!2\int \!d\phi\, (\sqrt{|1-\phi^2|})^{2N-1} \\
\!\!&=&\!\! 2\phi \times \,{_2}F_1\!\left(\frac12,-N+\frac12;\,\frac32;\,\phi^{2}\!\right)\!\!A(\phi) + B(\phi)\label{W_compact}\nonumber\een
where
\ben
A(\phi)&=&\begin{cases}  (-1)^{N} i,  &  \mbox{for} \,\,\,\phi^2>1\;\\ 
1, & \mbox{for} \,\,\,\,\phi^2\leq1 \end{cases}
\nonumber\\
B(\phi)&=&\begin{cases}  \displaystyle {\rm sgn}(\phi)(1\!-\!i(-1)^N) \frac{\sqrt{\pi}\,\Gamma(N\!+\!1/2)}{\Gamma(N+1)} &\!\!\!  \mbox{for}\;\phi^2>1\;\\ \displaystyle
0, &\!\!\! \mbox{for } \phi^2\leq1 \nonumber \end{cases}
\een
For example, we write explicitly 
\ben
W(\phi) \!\!&=&\!\! 
\begin{cases} \displaystyle
g_1(\phi)+ {\rm sgn}(\phi)\!\!\left(\arccosh(|\phi|)-\frac{\pi}{2} \right) &  \mbox{for}\, \phi^2>1\;\;\\ \displaystyle
g_1(\phi) + \arcsin(\phi) & \mbox{for} \,\phi^2\leq 1\;\;
\end{cases}\nonumber\\
W(\phi)\!\!&=&\!\! 
\begin{cases} 
\displaystyle g_2(\phi) + \frac{3{\rm sgn}(\phi)}{4}\!\left(\arccosh(|\phi|)+{2\pi} 
\right)&\!\! \! \mbox{ for} \,\phi^2>1\;\; \displaystyle \\ 
\displaystyle -g_2(\phi) + \frac34 \arcsin(\phi) &\!\! \!\mbox{ for}\, \phi^2\leq 1 \;\;\nonumber\displaystyle\end{cases}
\een
where 
\ben 
g_1(\phi)&=&\phi \sqrt{|1-\phi^2|} \\
g_2(\phi)&=&-\frac12\phi \sqrt{|1-\phi^2| }\left(\frac52 -\phi^2\right),
\een
 for $n=1$ and for $n=2$, respectively. The potential have three local extrema: one maximum $\phi_0$, at $\phi_0=0$, where $V(\phi_0)=(2n-1)/(2n)$ is always a positive constant. The two other are local minima, at $\phi_{\pm}=\pm 1$, with
\be
V(\phi_{\pm})=-\frac{\pi}3 \left(\frac{ \Gamma(n+\frac12)}{\Gamma(n+1)}\right)^2,
\ee
with negative value. The behavior of the $\phi_0$ and $\phi_{\pm}$ as a function on $n$ is shown in Fig. \ref{figure4}. Moreover, we can note that $V(\phi\to\pm\infty) \to -\infty$. 

\begin{figure}[h!] 
\centering
\includegraphics[scale=0.45]{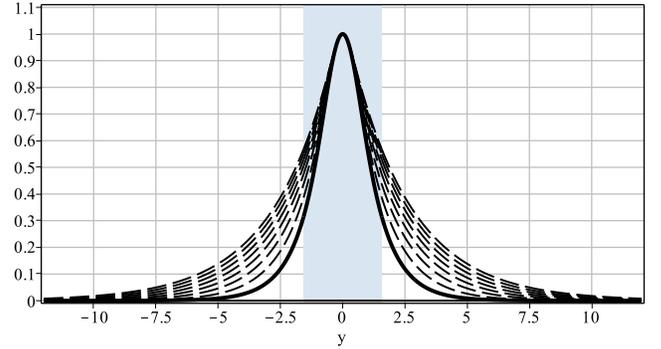} 
\caption{\label{figure5a}Profile of the warp factor $e^{2A(y)}$, for $n=1$ (solid line) and for $n=2,3,4,5,6,7,8$ (dashed lines). The shadow portion represents the region where the field is not constant ($|y|<\pi/2$).}
\end{figure}

The warp function can be written as
\ben
A(y) &=& \nonumber
\begin{cases} \displaystyle
-\frac{\pi}{6} |y|+\frac{\pi^2-4}{24}&  \mbox{ for } |y|>\frac{\pi}{2}\\ 
-\! \displaystyle\frac{y^2}6-\frac{\sin^2(y)}{6} & \mbox{ for } |y|\leq\frac{\pi}{2} 
\end{cases}\\
A(y) &=& 
\begin{cases} 
\displaystyle -\frac{\pi}{8} |y|-\frac{16-3\pi^2}{96}&  \mbox{ for } |y|>\frac{\pi}{2}\\ 
\displaystyle -\frac{y^2}8-\frac5{24}\sin^2(y)+\frac{1}{24}\sin^4(y) & \mbox{ for } |y|\leq\frac{\pi}{2}\end{cases} \nonumber
\een
for $n=1$ and for $n=2$, respectively. We fix $A(0)=0$. Note that for $|y|>\pi/2$, the behavior of the warp factor is similar to the case of  a thin brane. This happens because the scalar field is compact, so it is at a local minimum of the potential for $|y|>\pi/2$. In the Fig.~\eqref{figure5a}, we plot the profile of warp factor for some values of $n$.   

Similar behavior is also found for the corresponding energy densities, 
\ben
\rho(y)\!\!&=&\!\! \displaystyle
\begin{cases} \displaystyle
-\frac{\pi^2}{12}e^{-\frac{\pi}{3} |y|+\frac{\pi^2-4}{12}} &  \mbox{ for } |y|>\frac{\pi}{2}\\ \nonumber
f_1(y) e^{-\!\frac{y^2}3-\frac{\sin^2(y)}{3}}&  \mbox{ for } |y|\leq\frac{\pi}{2}\end{cases}
\\
\rho(y)\!\!&=&\!\! \displaystyle
\begin{cases} \displaystyle
-\frac{3\pi^2}{64}e^{-\frac{\pi}{4} |y|-\frac13+\frac{\pi^2}{16}} & \!\!\!\! \mbox{ for } |y|>\frac{\pi}{2}\\ 
f_2(y)e^{-\frac{y^2}4-\frac5{12}\sin^2(y)+\frac{1}{12}\sin^4(y)}& \!\!\!\!\mbox{ for } |y|\leq\frac{\pi}{2} \end{cases}\nonumber
\een
where
\ben
f_1(y)\!&=&\! 1-\frac{y^2}3 +\frac{y}3\sin(2y)-\frac4{3}\sin^2(y)+\frac{1}3\sin^4(y) \nonumber\\
f_2(y)\!&=&\!\frac12-\frac{3y^2}{16}-\frac{y}8\left(\frac{5}2-\sin^2(y)\right)\sin(2y)-\nonumber\frac{23}{48}\sin^2(y)\\&&-\frac7{16}\sin^4(y)+\frac12\sin^6(y)-\frac1{12}\sin^8(y) \nonumber
\een
for $n=1$ and for $n=2$, respectively. 

\begin{figure}[h!] 
\centering
\includegraphics[scale=0.45]{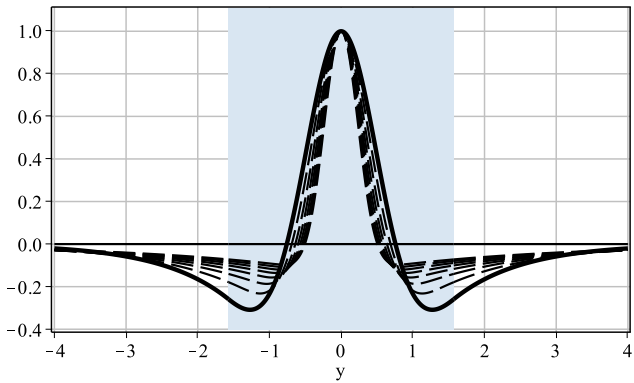} 
\caption{\label{figure6a} Profile of the energy density $\rho(y)$, for $n=1$ (solid line) and for $n=2,3,4,5,6,7,8$ (dashed lines). The shadow portion represents the region where the scalar field is not constant ($|y|<\pi/2$).} 
\end{figure} 

Before closing this section, let us study the interesting case where two distinct models lead to the very same warp function $A(y)$. 
The issue here is motivated by recent investigations on twinlike models, which are different models supporting the very same defect structure \cite{TM}.
The idea is to construct two distinct models, supporting different defect structures, but giving rise to the very same warp factor.
Using the Eq.~\eqref{Apij}, we have 
\be
A^{\prime\prime}=-\frac23 {\phi^{\prime2n }}
\ee
which can be related via two distance models, with different values of $n$ (say, $n_1$ and $n_2$). If we impose that the warp factor is the same, we get that
${\phi_1^\prime}^{n_1}={\phi_2^\prime}^{n_2}$. We illustrate this choosing $n_1=1$ and $n_2=2$ and the warp function \eqref{A1igual}. In this case, for the first model, with $n_1=1$, we get the solution given by Eq. \eqref{solkink}, where $\phi_1^\prime={\sech}^2(y)$. On the other hand, for the second model, with $n_2=2$ we get $\phi^{\prime}_2={\sech(y)}$, and the respective solution $\phi_2(y)={\rm arcsinh}(\tanh(x))$. Now, it is not hard to see that this solutions can be obtained with
\be
W(\phi)=\frac32 \sin(\phi) + \frac{1}{6} \sin(3\phi).
\ee
This ends the calculation. In particular, we note that our focus on the construction of the model from the defect structure the scalar field engenders, is crucial to implement the above issue, that allows obtaining different models that support the very same warp factor.

\subsection{Two-field models}

Let us now consider models described by two real scalar fields. We use the model \eqref{L.modifi} with $n=2$. In the case of one-field models, we compared 
models with distinct $n$. Here, however, we compare two possible solutions for the same $n$. 

Taking $\phi_1=\phi$ and $\phi_2=\chi$, we can write 
\begin{equation}\label{twofield1}
{\cal L}=X|X|-V(\phi,\chi),
\end{equation}
and assuming that $h_{ij}=\delta_{ij}$ we obtain
\begin{equation}\label{twofield2}
X=-\frac12\phi^{\prime 2}-\frac12\chi^{\prime 2}.
\end{equation}
Equation \eqref{akdapo} allows us to write
\ben\label{eqdiferen}
\phi^\prime |X|= \frac14 W_{\phi}(\phi,\chi) \,\,\,\, {\rm and} \,\,\,\, \chi^\prime |X|= \frac14 W_{\chi}(\phi,\chi).
\een
We use these equations to write \be\phi^{\prime 2}+\chi^{\prime 2}= \left(\frac14 W_{\phi}^2+\frac14 W_{\chi}^2\right)^{\frac13}.
\ee
Eq.~\eqref{alkl} can be used to write the potential as 
\ben
V(\phi,\chi)=\frac{3}{4}\left(\frac14 W_{\phi}^2+\frac14 W_{\chi}^2\right)^{\frac23}-\frac13 W^2.
\een
Now, we take the two Eqs.~\eqref{eqdiferen} to decouple the derivates of fields $\phi^\prime$ and $\chi^\prime$; we get   
\bes\label{eqsolutions1}
\ben
\phi^\prime &=& \frac1{2^{1/3}} \frac{W_{\phi}}{\left(W_{\phi}^2+W_{\chi}^2\right)^{1/3}},\label{eqp}\\
\chi^\prime &=& \frac1{2^{1/3}} \frac{W_{\chi}}{\left(W_{\phi}^2+W_{\chi}^2\right)^{1/3}}.\label{eqc}
\een
\label{eqsss}
\ees
We see that the set of constant and uniform solutions can be found taking $W_\phi=0$ and  $W_\chi=0$. Note that these solutions identify the local minima of the potential, and they make the potential vanish in Minkowski space, or be a negative constant in anti-de Sitter space.

The warp function $A(y)$ can be expressed as a function of the fields $\phi$ and $\chi$; using \eqref{aprime} we get 
\be\label{warpequation}
 {W_{\phi}A_{\phi}+W_{\chi}A_{\chi}}=-\frac{2^{1/3}}3 W {\left(W_{\phi}^2+W_{\chi}^2\right)^{1/3}}.
\ee
We can also use Eqs. \eqref{eqsolutions1} to get
\begin{equation}\label{eqorbit}
\frac{d\phi}{d\chi}=\frac{W_{\phi}}{W_{\chi}}.
\end{equation}
Solutions of this equation are orbits in the plane $(\phi,\chi)$.

To show how to solve this problem, let us consider the specific model
\begin{equation}
W(\phi,\chi)=2\phi-\frac23\phi^3-2r\phi\chi^2.
\end{equation}
This model was studied in several works, see e.,g., Ref.~\cite{BB}. In the present investigation, the potential has the form
\begin{eqnarray}
V(\phi,\chi)&=&\frac{3}{4}\left[(1-\phi^2-r\chi^2)^2+(2r\phi\chi)^2\right]^{\frac23}-\nonumber\\
&-&\frac43\left (\phi-\frac13\phi^3-r\phi\chi^2\right)^2.
\end{eqnarray}
The Eqs.~\eqref{eqsolutions1} can be written as
\bes\label{eqsolutions2}
\ben
\phi^\prime &=&\frac{1-\phi^2-r\chi^2}{\left[(1-\phi^2-r\chi^2)^2+(2r\phi\chi)^2\right]^{1/3}}.\label{eqp2}\\
\chi^\prime &=& - \frac{2r\phi\chi}{\left[(1-\phi^2-r\chi^2)^2+(2r\phi\chi)^2\right]^{1/3}}.\label{eqc2}
\een
\ees
There are four homogeneous solutions, for $r>0$: $v_1=(-1,0)$, $v_2=(0,1/\sqrt{r})$, $v_3=(1,0)$ and $v_4=(0,-1/\sqrt{r})$. Note that $V(v_2)=V(v_4)=0$, $V(v_1)=V(v_3)=-4/{27}$.
We use \eqref{eqorbit} to get
\begin{equation}\nonumber
\frac{d\phi}{d\chi}=-\frac{1-\phi^2-r\chi^2}{2r\phi\chi}.
\end{equation}
We use this equation to obtain the orbit
\begin{equation}\label{solorbit}
\phi^2=1+\frac{r}{2r-1}\chi^2+c\chi^{\frac1r},
\end{equation}
where $c$ is a real constant and $r\neq 1/2$. Each orbit leads to a couple of solutions $\phi(y)$ and $\chi(y)$; consequently, a different $A(y)$ is obtained case by case.

To illustrate, let us study two distinct cases, describing two distinct orbits.
Both orbits connect the two minima $v_1$ to $v_3$. Firstly, we get the orbit given by the straight line obtained by the condition $\chi=0$. In this case, the equation \eqref{eqp2} is written as
\be
\phi^\prime =(1-\phi^2)^{1/3},
\ee
and according to \cite{BL2011} has the folowing solution
\begin{equation}\label{compacttwofield}
\phi(y) = 
\begin{cases} 
{\rm sgn }(y) &  \mbox{ for } |y|>L_s,\\ 
\phi_s(y) &  \mbox{ for } |y|\leq L_s,
\end{cases}
\end{equation}
where $\phi_s(y)$ is the solution of the transcendental equation
\begin{equation}\label{sol2cam1}
\phi_s(y) \;{_2F_1}\left[\frac13,\frac12;\frac32;\phi_s(y)^2\right]=y,
\end{equation}
and $L_s=\displaystyle{\displaystyle3^{\frac32}2^{-\frac53} \Gamma^3({2}/{3})}\pi^{-1}\simeq 1.293$. Using the Eq. \eqref{warpequation}, we write the warp factor as
\begin{equation}\label{compacttwofield}
A_s(y) = 
\begin{cases} 
-\displaystyle\frac{4}{9}|y|+\frac25 +\frac49 L_s &  \mbox{ for } |y|>L_s,\\ 
A(\phi_s(y)) &  \mbox{ for } |y|\leq L_s,
\end{cases}
\end{equation}
where 
\begin{equation}
A(\phi_s)=\frac{1}{15}\left(1-\phi_s^2\right)^{2/3}\left(6-\phi_s^2\right)+\frac25,
\end{equation}
with $\phi_s(y)$ being a solution of Eq.~\eqref{sol2cam1}.

The second orbit is elliptical and can be obtained if we make $c=0$ in \eqref{solorbit}. In this case we obtain
\begin{equation}
\chi^2=\left(\frac1r-2\right)(1-\phi^2),
\end{equation}
for $0<r<1/2$. Substituting this orbit into Eq.~\eqref{eqp2}, we get
\begin{equation}
\phi^\prime =\frac{[2r(1-\phi^2)]^{2/3}}{\left[2r-2(3r-1)\phi^2\right]^{1/3}}.
\end{equation}

The solution depends on $r$. For simplicity, we choose the case $r=1/3$, and now the above equation becomes
\begin{equation}
\phi^\prime ={(2/3)^{1/3}(1-\phi^2)^{2/3}}.
\end{equation}
We obtain the solution as
\begin{equation}\label{compacttwofield}
\phi(y) = 
\begin{cases} 
{\rm sgn }(y) &  \mbox{ for } |y|>L_e,\\ 
\phi_e(y) &  \mbox{ for } |y|\leq L_e,
\end{cases}
\end{equation}
where $\phi_e$ is the solution of the transcendental equation
\ben\label{setetete}
\phi_e(y) \;{_2F_1}\left[\frac12,\frac23;\frac32;\phi_e(y)^2\right]\!=\!\!(2/3)^{1/3}y,
\een
and $L_e=\displaystyle{\displaystyle3^{-\frac23}2^{\frac13}\pi^2 }\Gamma^{-3}({2}/{3})\simeq 4.815$. 
 Using the Eq. \eqref{warpequation}, we write the warp factor as
\begin{equation}\label{compacttwofield}
A_e(y) = 
\begin{cases} 
-\displaystyle\frac{8}{9}|y|+ \frac{1}{2^{\frac13}3^{\frac23}} -\frac89 L_e &  \mbox{ for } |y|>L_e,\\ 
A(\phi_e(y)) &  \mbox{ for } |y|\leq L_e,
\end{cases}
\end{equation}
where 
\begin{eqnarray}
A(\phi_e)=(4/9)^{1/3}\left[(1-\phi_e^2)^{1/3}-1\right],
\end{eqnarray}
with $\phi_e(y)$ is solution of Eq. \eqref{setetete}. 

\begin{figure}[h!] 
\centering
\includegraphics[scale=0.45]{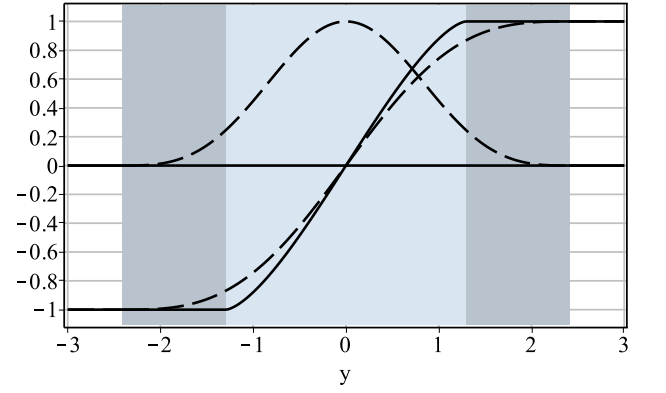} 
\caption{\label{figure7}Profile of the solutions $\phi(y)$ and $\chi(y)$ that obey Eqs.~\eqref{eqsss}, in the case of the straight line orbit (solid lines) and the elliptical orbit (dashed lines). Both solutions are compact, and in the figure the shadow portions represent regions where the fields are not constant, for $|y|<L_s$ and for $|y|<L_e$, respectively.} 
\end{figure}

The profile of the solutions $\phi(y)$ and $\chi(y)$ is shown in Fig.~\ref{figure7}. They are compact solutions for both the straight and elliptical orbits. For these solutions, the thickness is well defined, giving by $L_s$ and $L_e$. We note that $L_e/L_s\simeq 1.861$. 

\begin{figure}[h!] 
\centering
\includegraphics[scale=0.45]{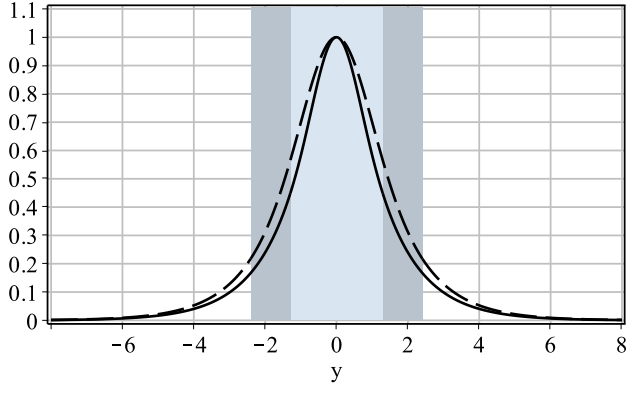} 
\caption{\label{figure8}
Profile of the warp factor in the case of the straight line orbit (solid lines) and the elliptical orbit (dashed lines). The shadow regions indicates where the scalar fields are not constant.} 
\end{figure}

In the absence of gravity, in the standard case, the solutions for the two distinct orbits are
\ben
\phi(y)=\tanh(y) \! &{\rm and}&\! {\chi(x)=0},\nonumber \\
\phi(y)=\tanh(2ry) \! &{\rm and}& \! \chi(x)=\sqrt{\frac{1-2r}{r}} \sech(2ry),\nonumber
\een 
respectively. Here, the ratio between the thickness of the two solutions is $1/2r$, and for $r=1/3$, it gives $1.5$. In the Fig. \ref{figure8}, we plot the profile of the warp factor
for the two solutions, obeying the straight and elliptical orbits. 

\section{Stability} \label{stable}

The present study is of direct interest to high energy physics, but it is important to know if the modifications introduced in the scalar field sector contribute to destabilize the geometric degrees of freedom of the braneworld model. To investigate this issue, let us study linear stability in the usual way. We consider
perturbations in the form
\begin{equation}\label{eq1.55}
\overline{g}_{ab}=g_{ab}+\pi_{ab}(y, \textit{x}).
\end{equation}
The perturbation $\pi_{ab}$ obeys the restriction 
$\pi_{\mu 4}(y,\textit{x})=0.$ Also, we have to have  $\pi^{ab}=-g^{am}\pi_{mn}g^{nb}$. We rewrite  $\pi_{\mu\nu}(y,\textit{x})$  as $\pi_{\mu\nu}(y,x)=e^{2{A(y)}}h_{\mu\nu}(y,\textit{x})$,
and now the perturbed line element has the form
\begin{equation}\label{eq1.58}
ds^2=e^{2A(y)}(\eta_{\mu\nu}+h_{\mu\nu}(y,\textit{x}))dx^\mu dx^\nu-dy^2.
\end{equation}

We must also consider fluctuations on the set of scalar fields  
\begin{equation}\label{eq1.60}
\overline{\phi}_i=\phi_i(y)+\xi_i(y,x).
\end{equation}
The first-order contribution to the fluctuations of the term $X_{ij}$ is written as
\be
X^{(1)}_{ij}=\frac12\!\left( \nabla_a\phi_i\nabla^a\xi_j\!+\! \nabla_a\xi_i\nabla^a\phi_j\!+\!\pi_{ab}\nabla^a\phi_i\nabla^b \phi_j \right).
\ee 
Also, the first-order contribution of the Einstein equations in Ricci tensor appears as $R_{ab}=2\bar{T}_{ab}$, with $\bar{T}_{ab}={T}_{ab}-\frac13 g_{ab} T^c_{\,\,c},$ and 
\begin{subequations}\label{eq1.67}
\begin{eqnarray}
\bar{T}_{\mu\nu}^{(1)}&=& \nonumber\frac23e^{2{ A}} {\eta}_{\mu\nu}\Big[-X_{ij}\left({\cal L}_{X_{ij}\phi_k} \xi_k-{\cal L}_{X_{ij}X_{kl}}\phi_k^{\prime}\xi_l^{\prime}\right),
\\&&+{\cal L}_{\phi_k}\xi_k\Big]-\frac23 e^{2{ A}}h_{\mu\nu}\left(X_{ij}{\cal L}_{X_{ij}}-{\cal L}\right),\label{eq1.67.a}\\
\bar{T}_{\mu 4}^{(1)}&=&\phi_j^{\prime}{\cal L}_{X_{ij}}\nabla_\mu\xi_i,\label{eq1.67.b}\\
\bar{T}_{44}^{(1)}&=&- \nonumber\frac23\left(2X_{ij}{\cal L}_{X_{ij}\phi_k}+{\cal L}_{\phi_k}\right) \xi_k+\\&& \frac23\left(2X_{ij}{\cal L}_{X_{ij}X_{kl}}+3{\cal L}_{X_{kl}}\right)\phi_k^{\prime}\xi_l^{\prime},\label{eq1.67.c}
\end{eqnarray}
\end{subequations}
where ${\eta}_{\alpha\beta}$ is the metric on the Minkowski space.

Thus, Einstein's equation can be written in components. The $\{\mu,\nu\}$-component becomes:
\begin{subequations}\label{eq1.77}
\ben
&&\!\!\!\!\!\! e^{2{A}}\left(\frac12\partial_y^2+2{A}^{\prime}\partial_y\right)h_{\mu\nu}+\frac{1}{2}{\eta}_{\mu\nu}e^{2{A}}{A}^{\prime}\partial_y\left(\eta^{\alpha\beta} h_{\alpha\beta}\right)\,\,\,\,\,\\&&\!\!\!\!\!\! -\frac{1}{2}\eta^{\alpha\beta}\left(\partial_\alpha\partial_\beta h_{\mu\nu}\nonumber-\partial_\mu\partial_\nu h_{\alpha\beta}+\partial_\mu\partial_\alpha h_{\nu \beta}+\partial_\nu\partial_\alpha h_{\mu\beta}\right)\nonumber\\
&&\!\!\!\!\!\!=\frac43e^{2{A}} \eta_{\mu\nu}\left[-X_{ij}\left({\cal L}_{X_{ij}\phi_k} \xi_k-{\cal L}_{X_{ij}X_{kl}}\phi_k^{\prime}\xi_l^{\prime}\right)+{\cal L}_{\phi_k}\xi_k\right];\label{eq1.77.a}\nonumber
\een
the $\{\mu,4\}$-component is
\be
\frac{1}{2}\eta^{\nu \sigma}\partial_y\left(\partial_\nu h_{\sigma\mu}-\partial_\mu h_{\nu \sigma}\right)=2\phi_j^{\prime}{\cal L}_{X_{ij}}\nabla_\mu\xi_i;\label{eq1.77.b}
\ee
and finally the $\{4,4\}$-component has the form 
\ben
&&-\frac{1}{2}\left(\partial_y^2+2{A}^{\prime}\partial_y\right) (\eta^{\alpha\beta} h_{\alpha\beta})=\nonumber\\&&-\frac43\left(2X_{ij}{\cal L}_{X_{ij}\phi_k}+{\cal L}_{\phi_k}\right) \xi_k+\nonumber\\
&&+\frac43\left(2X_{ij}{\cal L}_{X_{ij}X_{kl}}+3{\cal L}_{X_{kl}}\right)\phi_k^{\prime}\xi_l^{\prime}. \label{eq1.77.c}
\een
\end{subequations}
The equation of motion for the scalar field gives
\begin{eqnarray}
&&{\cal L}_{X_{ij}}e^{-2{A}}\square\xi_j-\left[\left(2X_{kj}{\cal L}_{X_{ij}X_{kl}}+{\cal L}_{X_{il}}\right)\xi_l^{\prime}\right]^{\prime}\nonumber
\\
&&
-4{A}^{\prime}\left(2X_{kj}{\cal L}_{X_{ij}X_{kl}}+{\cal L}_{X_{il}}\right)\xi_l^{\prime}\nonumber\\
&&-\left[4{\cal L}_{ X_{ij}\phi_k}\phi_j^{\prime}{A}^{\prime}+\left({\cal L}_{ X_{ij}\phi_k}\phi_j^{\prime}\right)^{\prime}+{\cal L}_{\phi_i\phi_k}\right]\xi_k
\nonumber\\
&
&+\left({\cal L}_{\phi_i X_{jk}}-{\cal L}_{\phi_k X_{ij}}\right)\phi_j^{\prime}\xi_k^{\prime}\nonumber\\
&=&\frac12{\cal L}_{X_{ij}}\phi_j^{\prime}\eta^{\alpha\beta}
h_{\alpha\beta}^{\prime}.
\end{eqnarray}

Let us now consider the transverse traceless components for metric  fluctuations
\begin{equation}\label{eq1.80}
\overline{h}_{\mu\nu}=\left(\frac12(\tau_{\mu \alpha}\tau_{\nu \beta}+\tau_{\mu \beta}\tau_{\nu \alpha})-\frac13\tau_{\mu\nu}\tau_{\alpha\beta}\right)h^{\alpha\beta},
\end{equation}
where $\tau_{\mu\nu}\equiv h_{\mu\nu}-\partial_\mu\partial_\nu/\square$. We note that the net effect of
this projection operation is to decouple the metric fluctuation equation from the scalar field equation, even in
the general case which is being considered in the present work. As a matter of fact, we can check that
\begin{equation}\label{eq1.83}
\left(\partial_y^2+4{ A}^{\prime}\partial_y-e^{-2{A}}\square\right)\overline{h}_{\mu\nu}=0.
\end{equation}
The next steps follow the standard procedure: we chance the $y$-coordinate for a $z$-coordinate, in order to make the metric conformally flat,
with $dz = e^{-A(y)}dy$. This allows changing the Eq.~\eqref{eq1.58} to
\begin{equation}\label{eq1.85}
ds^2=e^{2A(z)}\left[(\eta_{\mu\nu}+h_{\mu\nu}(z,\textit{x}))dx^\mu dx^\nu-dz^2\right],
\end{equation}
and now we can rewrite the Eq.~\eqref{eq1.83} as
$\left(-\partial_z^2-3{ A}_z\partial_z+\square\right)\overline{h}_{\mu\nu}=0$.
In order to remove the first derivative in this equation, we redefine the gravitational field $\overline{h}_{\mu\nu}$ in the following way: $\overline{h}_{\mu\nu}(\textit{x},z)\equiv e^{-3{ A}(z)/2}H_{\mu\nu}(\textit{x},z)$. This transforms the equation to
\begin{equation}\label{eq1.88}
\left(-\partial_z^2+U(z)+\square\right)H_{\mu\nu}=0,
\end{equation}
where
\begin{equation}\label{eq1.89}
U(z)=\frac94 A_z^2+\frac32 A_{zz}.
\end{equation}

We can make the following separation of variables: $H_{\mu\nu}({x},z)=\Psi({x})\tilde{H}_{\mu\nu}(z)$, where $\Psi({x})$ obeys the plane wave equation $\square \Psi({x})=-p^2\Psi({x})$. Thus, the above Eq.\eqref{eq1.88} changes to 
\begin{equation}\label{eq1.90}
\left(-\partial_z^2+U(z)\right)H_{\mu\nu}=p^2H_{\mu\nu}
\end{equation}
Note that we can rewrite this equation as 
\begin{equation}
{\cal Q}^{\dagger}{\cal Q}H_{\mu\nu}\!=\!\left[\left(\partial_z+\frac32{A_z}\right)\left(-\partial_z+\frac32{A_z}\right)\right]H_{\mu\nu}=p^2H_{\mu\nu}\nonumber
\end{equation}
This factorization directly shows that there are no graviton bound states with negative mass. The graviton
zero mode $H_{µν}(z) \propto e^{\frac32A(z)}$ is the ground-state of the
associated quantum mechanical problem. This leads to the important conclusion that the modification appearing from the $N$ scalar fields dynamics does not contribute to destabilize the geometric degrees of freedom which appears in the standard braneworld scenario. Thus, the proposed modification is robust.

\begin{figure}[h!] 
\centering
\includegraphics[scale=0.45]{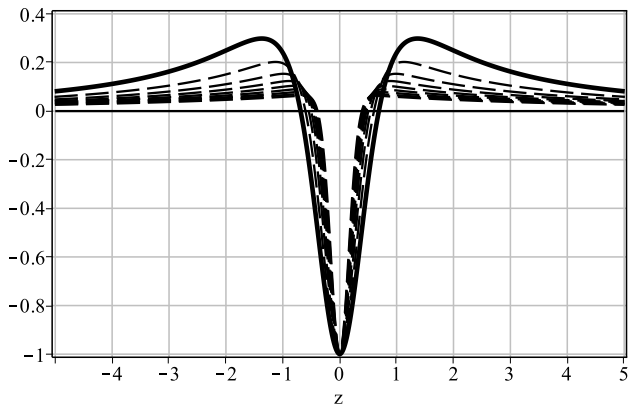} 
\caption{Profile of the Schroedinger-like potential \label{figure9} for the model given by $W$ as in \eqref{W_tanh}. The solid line refer $n=1$ while, the dashed lines refer to $n=2,3,...,8$. } 
\end{figure}

\begin{figure}[h!] 
\centering
\includegraphics[scale=0.45]{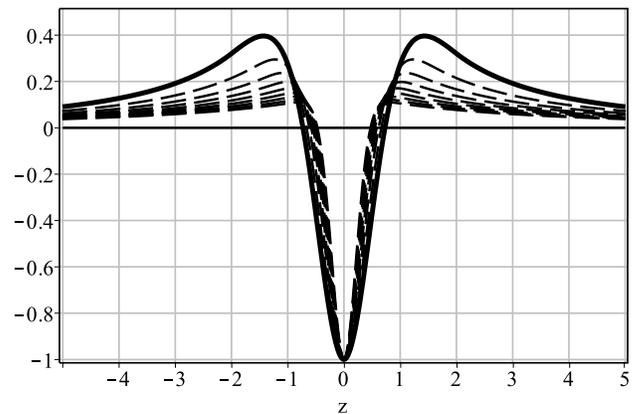} 
\caption{Profile of the Schroedinger-like potentials \label{figure10} for the model given by $W$ as in \eqref{W_compact}. The solid line refer $n=1$, while the dashed lines refer to $n=2,3,...,8$.} 
\end{figure}

For example, we plot the Schroedinger-like potentials for the models given by $W$ functions \eqref{W_tanh} and \eqref{W_compact}, for the standard kink (Fig.~\ref{figure9}) and compact solutions (Fig.~\ref{figure10}), respectively. We make the transformations of the variables $y$ to $z$ which depends on the warp functions $A(y)$.  In both cases, all potentials are vulcano-like, with their vibrational modes being asymptotically plane waves, with $k^2>0$. We note that the height of the maxima of the quantum-mechanical potential decreases for increasing $n$. 

\section{conclusions}\label{end}
 In this work we studied models described by $N$ real scalar fields, coupled to gravity in the brane scenario, with a single extra dimension of infinite extent.
 The main novelty of the investigation concerns the nonstandard dynamics that drives the scalar fields, and the focus on the construction of the model from the defect structure
 that solves the the first-order equations associated to the set of scalar fields. We illustrated the general results with several examples described by one and by two scalar fields, for several distinct generalized dynamics, governed by the integer $n$, as in Eq.~{\eqref{L.modifi}} and in Eqs.~{\eqref{twofield1}} and \eqref{twofield2}. 
 
 In the case of a single field, we also considered an interesting case, where we constructed the very same warp factor, using two distinct models, supporting distinct defect structures.  

In order to complete the investigation, we studied stability in the standard sense, introducing fluctuations in both the scalar fields and in the metric. We used explicit results to show that the fluctuations decouple, even though we are working in a more general scenario, where the dynamics of the scalar fields is changed to allow for higher order terms in the derivative of the fields.  In particular, we depicted the quantum mechanical potential associated to the fluctuations in the metric, in the case of two distinct situations, described by \eqref{W_tanh} and by \eqref{W_compact}, for several values on $n$. The results show that fluctuations in the metric follow the standard scenario, despite the generalized dynamics engendered by the scalar fields.

We would like to thank CAPES, CNPq and  FAPESP for partial financial support.


\end{document}